\documentclass[prb,twocolumn,showpacs,superscriptaddress]{revtex4}

\usepackage{graphicx}
\usepackage{dcolumn}
\usepackage{amsmath}
\usepackage{color}

\newcommand{\SrCrO}{Sr$_{3}$Cr$_2$O$_8$}
\newcommand{\BaCrO}{Ba$_{3}$Cr$_2$O$_8$}

\begin{document}
\title{Infrared phonons and specific heat in the gapped quantum magnet Ba$_{3}$Cr$_2$O$_8$}

\author{Zhe~Wang}
\author{M.~Schmidt}
\author{A.~G\"{u}nther}
\author{F.~Mayr}
\affiliation{Experimental Physics V, Center for Electronic
Correlations and Magnetism, Institute of Physics, University of Augsburg, D-86135 Augsburg, Germany}

\author{Yuan~Wan}
\affiliation{Department of Physics and Astronomy, Johns Hopkins University, Baltimore, Maryland 21218, USA}

\author{S.-H.~Lee}
\affiliation{Department of Physics, University of Virginia, Charlottesville, Virginia 22904, USA}

\author{H.~Ueda}
\author{Y.~Ueda}
\affiliation{Institute for Solid State Physics, University of Tokyo, Kashiwa 277-8581, Japan}

\author{A.~Loidl}
\author{J.~Deisenhofer}
\affiliation{Experimental Physics V, Center for Electronic
Correlations and Magnetism, Institute of Physics, University of Augsburg, D-86135 Augsburg, Germany}

\date{\today}

\begin{abstract}
We report on the phonon spectrum of the spin-gapped quantum magnet \BaCrO~determined by infrared
spectroscopy, and on specific heat measurements across the
Jahn-Teller transition in magnetic fields up to 9~T. Phonon modes
split below the Jahn-Teller transition, which occurs at $T_{JT}$ =
70~K as detected by specific-heat measurements. The field-dependent
specific heat data is analyzed in terms of the contributions from
lattice, magnetic and orbital degrees of freedom. In contrast to the
isostructural compound \SrCrO~our analysis does not indicate the
existence of orbital fluctuations below the Jahn-Teller transition
in \BaCrO.
\end{abstract}



\maketitle

\section{Introduction}

Spin-gapped quantum antiferromagnets with dimerized ground states  have been
intensively studied due to the observation of various exotic
phenomena in these systems.\cite{Lemmens03,Giamarchi08,Zakharov06,Wang11b}
Based on dimers of Cr$^{5+}$ (3d$^1$, \emph{S} = 1/2) or Mn$^{5+}$
(3d$^2$, \emph{S} = 1) ions, a series of isostructural dimerized
compounds, such as \BaCrO,\cite{Kofu09a,Aczel09PRB,Dodds10} \SrCrO,\cite{Aczel09}
and Ba$_{3}$Mn$_2$O$_8$\cite{Samulon09} exhibits a
magnon-condensation state with the application of an external
magnetic field.\cite{Giamarchi08}

The room-temperature crystalline structure of these compounds is
hexagonal with space group
\emph{R}$\bar{3}$\emph{m}.\cite{Kofu09a,Chapon08,Stone08} In
\BaCrO~and \SrCrO, a tetrahedral crystal field splits the 3\emph{d}
levels of the Cr ions into lower-lying doubly degenerated \emph{e}
orbitals and higher-lying triply degenerated \emph{t} orbitals.
Thus, the systems are Jahn-Teller active and they are expected to
undergo a structural phase transition to remove the orbital
degeneracy. Neutron powder diffraction indeed revealed such a
structural transition at 70~K in \BaCrO~and at 275~K in \SrCrO~from
hexagonal to monoclinic with space group
\emph{C}2/\emph{c}.\cite{Kofu09a,Chapon08}

The magnetic susceptibility of \BaCrO~has been reported by  Nakajima
\emph{et al.},\cite{Nakajima06} and Aczel \emph{et
al}.\cite{Aczel08} They have shown that the susceptibility can be
described by a Bleaney-Bowers model,\cite{B.B.} where they
considered a magnetic structure consisting of weakly-interacting
spin dimers of Cr$^{5+}$ ions. Fitting to this model resulted in an
intra-dimer exchange interaction $J_0=2.15$~meV consistent with the
value of 2.38~meV determined by neutron experiments.\cite{Kofu09a}
The detailed magnetic structure has been studied theoretically based
on extended H\"{u}ckel tight binding calculations.\cite{Koo06} The
authors of Ref. \onlinecite{Koo06} have examined all the possible
exchange paths between the nearest-neighboring Cr ions. They found
that the exchange interaction between adjacent Cr$^{5+}$ ions in the
hexagonal $c_h$ direction is most important, while the other
exchange interactions are much smaller. This was confirmed by
inelastic neutron experiments in \BaCrO.\cite{Kofu09a} The magnetic
structure of \SrCrO~is the same as \BaCrO, but the intradimer
exchange interaction of \SrCrO~is 5.55~meV,\cite{Castro10} about twice the corresponding value in
\BaCrO, and the Jahn-Teller transition temperature 285~K in \SrCrO~is about
four times higher than that in \BaCrO.\cite{Wang11} In addition, an anomalous strong damping of the optical
phonon modes has been observed down to about
120~K in \SrCrO, which is below the Jahn-Teller transition
temperature of \SrCrO.\cite{Wang11,Wulferding11} Moreover, an analysis of the
specific heat revealed a residual orbital entropy below the
structural transition, indicating strong fluctuations in the orbital
degrees of freedom.

In this work we investigated the polar phonons and the specific heat
in \BaCrO~to search for a similar fluctuation regime. We have
resolved the phonon spectra from the reflectivity measurements and
found the splitting of phonon modes right below the Jahn-Teller
transition, indicating that strong fluctuations of the lattice
degrees of freedom are absent in \BaCrO. This clearly distinguishes the system from \SrCrO.
The analysis of the specific heat confirms this difference between the compounds, because the orbital entropy associated with the splitting of the orbital ground state is completely recovered at the Jahn-Teller transition in \BaCrO.

\section{Experimental details}\label{Sec:ExperDetail}
Single crystals of \BaCrO~were grown by the floating-zone method.
They were cleaved perpendicular to
the hexagonal $c_h$-axis and checked by single-crystal x-ray diffraction.
The heat capacity was measured in a
Quantum Design physical properties measurement system at
temperatures 1.8~$<T<$~300~K with applied magnetic fields up to 9~T.
Reflectivity measurements were carried out on optically polished samples
for 4~$<T<$~300~K in the
far- and mid-infrared range using the Bruker Fourier-transform infrared
spectrometers IFS 113v and IFS 66v/S with a He-flow cryostat
(Cryovac).

\section{Experimental results and discussion}
\subsection{Infrared spectra}\label{Sec:InfraredMode}

\begin{figure}[b]
\centering
\includegraphics[width=80mm,clip]{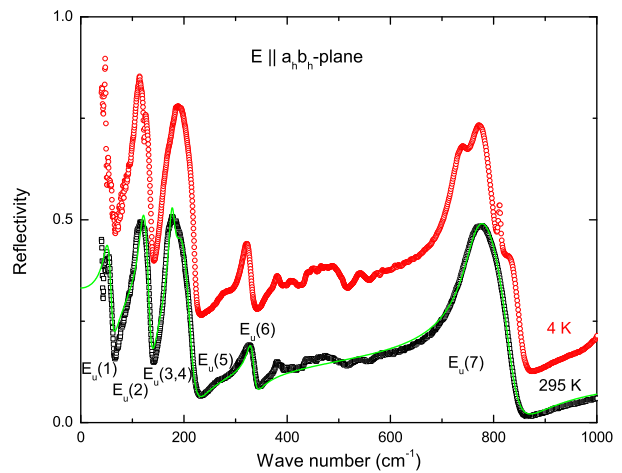}
\vspace{2mm} \caption[]{\label{Fig:IR_phonon} (Color
online)  Reflectivity spectra of \BaCrO~measured with $\mathbf{E}$ $\|$
$a_{h}b_{h}$-plane at 295~K and 4~K.
The spectrum of 4~K is shifted with respect to that of 295~K by a constant of 0.1.
Infrared-active phonon mode $E_u(j) (j=1,...,7)$ are marked at the spectrum of 295~K for the hexagonal structure.
Solid line is a fit of Drude-Lorentz model to the data of 295~K as described in the text.}
\end{figure}

Figure~\ref{Fig:IR_phonon} shows reflectivity spectra of \BaCrO~measured for the radiation
electric field $\mathbf{E}$ parallel with the crystalline hexagonal $a_hb_h$-plane ($\mathbf{E}\parallel a_hb_h$-plane)
at 295~K and 4~K, which are above and below the Jahn-Teller transition temperature $T_{JT}$ = 70~K, respectively.
The seven modes observed above $T_{JT}$ are tentatively assigned as $E_u$(\emph{j}) (\emph{j}=1,...,7). The primitive unit cell of the high-temperature hexagonal structure with space group $R\overline{3}m$ (No.~166)\cite{BuschbaumAczel,Hahn89} contains one chemical formula unit and, hence, a total of 39 normal modes exist. The IR active modes are characterized by the irreducible representations $6A_{2u}
(\mathbf{E}\parallel c_h)+7E_{u}[\mathbf{E}\parallel (a_h,b_h)]$.\cite{Kofu09a,
Kroumova} The number of normal modes is in agreement with the observation of infrared and Raman active phonons reported for the isostructural system \SrCrO.\cite{Wang11,Wulferding11} The overall appearance of the \BaCrO~far-infrared spectra are similar to the ones of \SrCrO, but the \BaCrO~phonons are shifted to lower eigenfrequencies consistent with the larger molar mass of \BaCrO.\cite{Wang11} The absolute value of the reflectivity is relatively low, which might be due to the imperfection of the sample surface, and similar values have been found for \SrCrO.\cite{Wang11} The weak features in the frequency range between modes $E_u$(6) and $E_u$(7) are also assigned to originate from defects on the sample's surface.

The quantitative analysis of the phonon modes is performed by fitting the reflectivity spectrum with the Drude-Lorentz model. According to this model, the phonon contribution of the complex dielectric function is given by
\begin{equation}\label{eq:DrudeLorentz}
\epsilon(\omega)=\epsilon_\infty+\sum_j\frac{\Omega_j^2}{\omega_{0j}^2-\omega^2-i\gamma_j\omega},
\end{equation}
where each phonon mode \emph{j} is described by the eigenfrequency
$\omega_{0j}$, damping coefficient $\gamma_j$ and effective ionic
plasma frequency $\Omega_j$. The parameter $\epsilon_\infty$ takes into account contributions of higher-lying electronic excitations and was used as a free fitting parameter ranging between $\sim4$ at room temperature and $\sim7$ just above the Jahn-Teller transition.  At normal incidence, the reflectivity
$R(\omega)$ is related to $\epsilon(\omega)$ via
\begin{equation}\label{eq:RandEps}
R(\omega)=\left |\frac{\sqrt{\epsilon(\omega)}-1}{\sqrt{\epsilon(\omega)}+1}\right|^2.
\end{equation}
The fit result for the room-temperature spectrum using seven oscillators is given by the solid line in Fig.~\ref{Fig:IR_phonon}, which agrees well with the experimental data. Note that the lowest lying mode $E_u$(1) could not be completely resolved and only the high frequency flank has been fitted. Thus, the fitting parameters for $E_u$(1) contain a larger experimental uncertainty. The fitting parameters for the 7$E_u$ modes are listed in Table~\ref{Tab:Fit200}.

\begin{table}[h]
\caption{\label{Tab:Fit200}  Eigenfrequency $\omega_0$, effective ionic plasma frequency $\Omega$ and damping coefficient $\gamma$ of all 7$E_u$ phonon modes in \BaCrO~at 295~K.}
\begin{ruledtabular}\vspace{0.5mm}
\begin{tabular}{c@{\hspace{3em}}*{3}{c}}
$E_u$  &\multicolumn{3}{c}{7 Phonons}
\\\hline
Mode &\multicolumn{1}{c}{$\omega_0$ (cm$^{-1}$)}  &\multicolumn{1}{c}{$\Omega$ (cm$^{-1}$)}  &\multicolumn{1}{c}{$\gamma$(cm$^{-1}$)}
\vspace{0.5mm}\\\hline
1  &52.6    &\multicolumn{1}{c}{101.2}  &\multicolumn{1}{c}{13.5}
\\
2  &119.7   &\multicolumn{1}{c}{187.9}  &\multicolumn{1}{c}{14.0}
\\
3  &174.7   &182.8       &10.6
\\
4  &188.4   &174.1       &26.2
\\
5  &274.7   &166.0       &172.1
\\
6  &330.5   &120.4       &14.9
\\
7  &759.2   &718.2       &45.5

\end{tabular}
\end{ruledtabular}
\end{table}

\begin{figure}[t]
\centering
\includegraphics[width=80mm,clip]{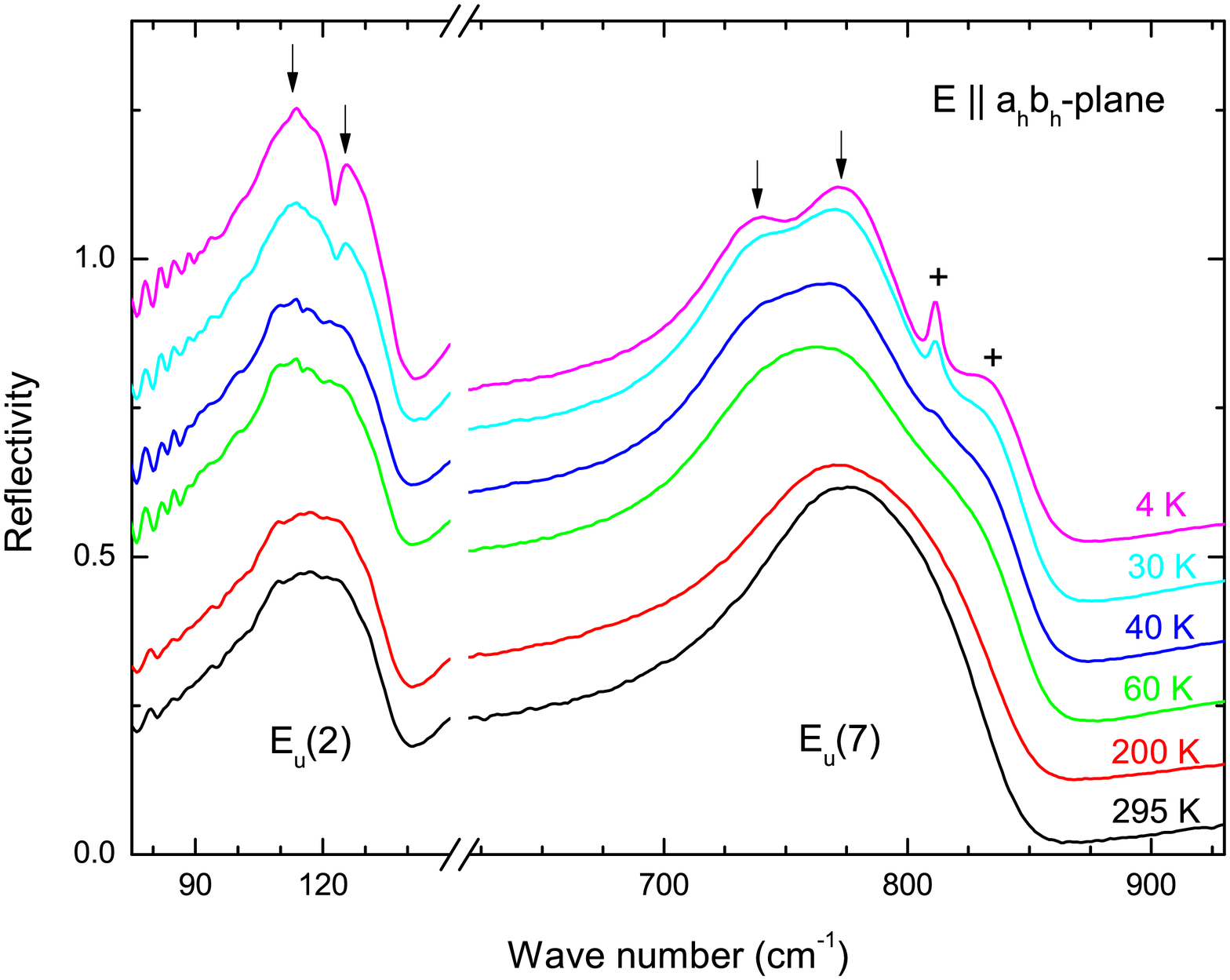}
\vspace{2mm} \caption[]{\label{Fig:EoneAndEseven} (Color
online) Reflectivity spectra corresponding to IR phonon modes
$E_u$(2) and $E_u$(7) measured with $\mathbf{E}$ $\|$ $a_{h}b_{h}$-plane at various temperatures.
The spectra are shifted with respect to that at 295~K by a constant in
order to clearly illustrate the evolution of phonon modes with decreasing temperature.}
\end{figure}

For the monoclinic $C2/c$ (No.\,15b) structure below the JT transition, the number of expected normal IR
modes increases to $19A_{u}$ for the radiation electric field parallel with monoclinic $b_m$ axis and
$20B_{u}$ for the electric field parallel with monoclinic $a_mc_m$-plane,
i.e. $\Gamma=19A_{u} (\mathbf{E}\parallel b_m)+20B_{u}[\mathbf{E}\parallel (a_m,c_m)]$.\cite{Kofu09a,Kroumova}
According to the geometric relation between hexagonal and monoclinic axes
$a_h=\frac{1}{2}(a_m-b_m)$, $b_h=-\frac{1}{2}(a_m+b_m)$, and $c_h=\frac{3}{2}c_m-\frac{1}{2}a_m$
determined by neutron diffraction measurements,\cite{Chapon08,Castro10}
the IR spectra measured with $\mathbf{E}$ $\|$ $a_{h}b_{h}$-plane below $T_{JT}$ should
display all the $19A_{u}$ and $20B_{u}$ modes. However, the coexistence of three monoclinic twins
in the single crystal\cite{Castro10,Kofu09a} and the low monoclinic symmetry make it difficult
to determine the eigenfrequency, damping coefficient and plasma frequency of the phonon modes below the Jahn-Teller transition.\cite{Kuzmenko96}

The temperature-dependent reflectivity spectra reveal that the hexagonal mode of $E_u$(2) and $E_u$(7) split below $T_{JT}$, while a splitting of the other modes, which are hardening with decreasing temperature, could not be resolved below $T_{JT}$. Therefore, we focus on the temperature-dependent features of modes $E_u$(2) and $E_u$(7), and try to describe the low-temperature modes with Lorentzian functions according to Eqs.~(\ref{eq:DrudeLorentz}) and (\ref{eq:RandEps}). As indicated by the arrows in Fig.~\ref{Fig:EoneAndEseven}, we model the low-temperature spectra with two oscillators for both the $E_u$(2) and the $E_u$(7) modes. For the monoclinic structure, the phonon modes cannot be uniquely determined in a single polarization measurement.
Therefore, the obtained fit parameters including $\epsilon_\infty$ do not necessarily reflect the intrinsic dielectric properties of the system and it might be misleading to show the dielectric parameters.
Instead, the reflectivity spectra should be measured with three different polarizations so as to determine the three independent matrix elements of the dielectric matrix for $\mathbf{E}\parallel a_mc_m$-plane.\cite{Kuzmenko96} This is infeasible for \BaCrO~with the coexistencce of three monoclinic twins.\cite{Castro10,Kofu09a} Thus the modeling here with Lorentzian functions is rather a way to parametrize the reflectivity spectra and reveal clear changes occurring at $T_{JT}$.

\begin{figure}[t]
\centering
\includegraphics[width=90mm,clip]{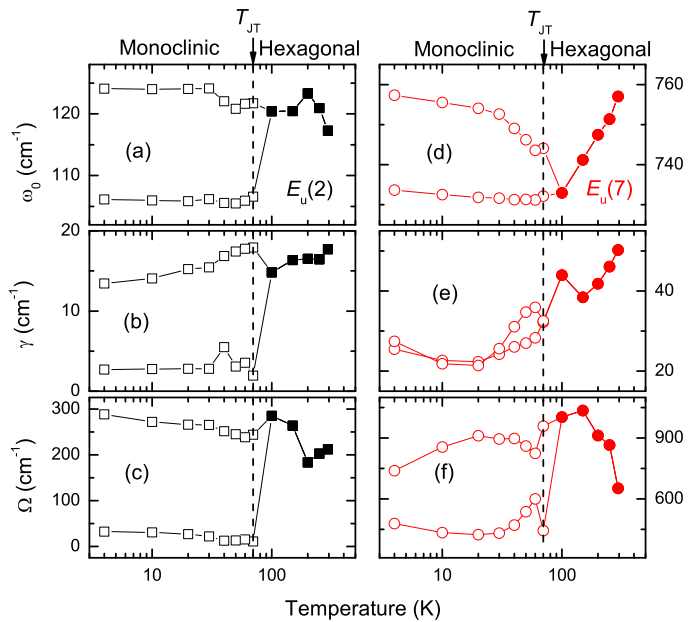}
\vspace{2mm} \caption[]{\label{Fig:EoneAndEsevenTDP} (Color
online) Temperature dependence of eigenfrequency (a, d), damping coefficient (b, e) and ionic plasma frequency (c, f)
for modes $E_u$(2) and $E_u$(7) on a semi-logarithmic scale.
The Jahn-Teller transition temperature $T_{JT}$ is indicated by vertical dashed line. Open symbols signify the results of parametrization below $T_{JT}$ (see text).}
\end{figure}

The fitting results of eigenfrequency, damping coefficient and plasma frequency are summarized as a function of temperature in Fig.~\ref{Fig:EoneAndEsevenTDP}, which clearly shows the splitting of phonons right below $T_{JT}$, consistent with the phase transition to monoclinic structure. This is an evident contrast to the case of \SrCrO, which is an iso-structural compound of \BaCrO.\cite{Chapon08} In \SrCrO, the new phonons are not emerging right below the Jahn-Teller transition due to strong fluctuations. With decreasing temperature, the fluctuations are reduced and all the modes of monoclinic structure are observed when the temperature is lower than 100~K.\cite{Wang11}

The modes marked by "+" in Fig.~\ref{Fig:EoneAndEseven} at 812~cm$^{-1}$ and 834~cm$^{-1}$ are visible right at 40~K and lower temperatures. These modes could be new phonon modes of the monoclinic structure, which have small spectral weight and large damping effect that could not be distinguished from the strong modes corresponding to $E_u$(7) at higher temperatures.

\subsection{Specific heat}\label{Sec:SpecHeat}

\begin{figure}[t]
\centering
\includegraphics[width=80mm,clip]{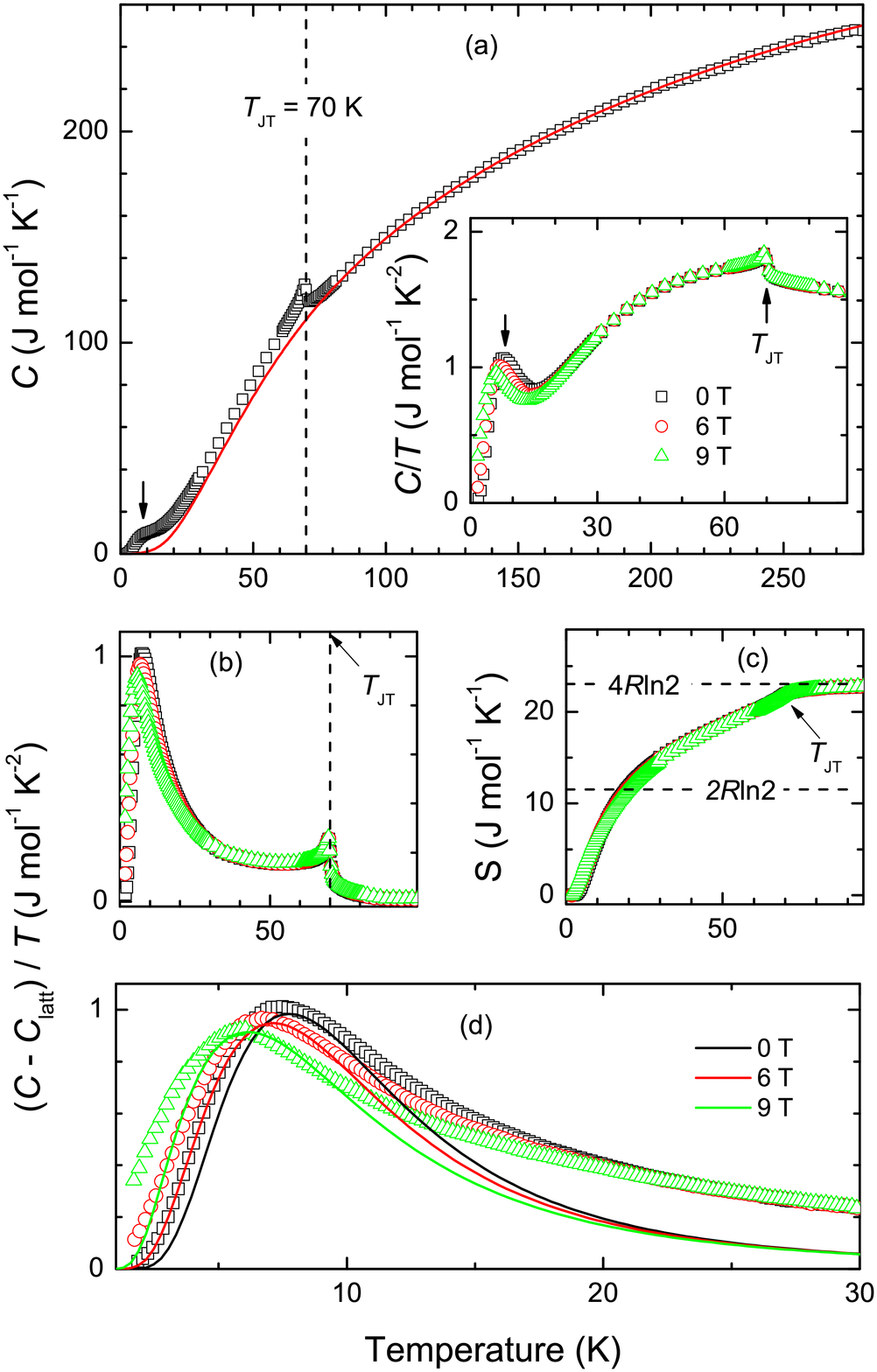}
\vspace{2mm} \caption[]{\label{Fig:SpecificHeat} (Color online)
(a) Specific heat of \BaCrO~without applied magnetic field. Solid line is a fit of lattice contribution to specific heat according to Debye and Einstein models as described in the text.
Inset shows specific heat divided by temperature $C/T$ measured at various magnetic fields. (b) Residual specific heat $(C-C_{latt})/T$ and (c) the corresponding entropy obtained at different fields. (d) Magnetic contribution to specific heat fitted by considering the Zeeman interaction of isolated spin dimers with magnetic field (solid lines) as described in the text.}

\end{figure}

Figure~\ref{Fig:SpecificHeat}(a) shows the specific heat of \BaCrO~ measured as a function of temperature. Two anomalies can be observed in the specific-heat data. A sharp anomaly indicated by a dashed vertical line can be easily recognized at 70~K. Another feature observed below 20~K is marked by an arrow. The inset of Fig.~\ref{Fig:SpecificHeat}(a) shows the specific heat over temperature $C/T$ vs. $T$ measured in various external magnetic fields. The anomaly at $T_{JT}$ = 70~K is independent on the magnitude of magnetic field, and marks the structural phase transition from hexagonal \emph{R}$\bar{3}$\emph{m} to monoclinic \emph{C}2/\emph{c}, in agreement with the results obtained from neutron diffraction measurements.\cite{Kofu09a} The anomaly below 20~K is very sensitive to the magnitude of the magnetic field indicating that is related to the spin degrees of freedom.

The lattice contribution to the specific heat $C_{latt}$ can be described according to the phononic spectra that have
been determined in Section \ref{Sec:InfraredMode}. We model the lattice contribution by a sum of one isotropic
Debye ($D$) and four isotropic Einstein terms ($E_{1,2,3,4}$) with the fixed ratio $D:E_1 :E_2
:E_3:E_4=1:3:4:3:2$ between these terms accounting for the 39 degrees of freedom per
formula unit. This ratio has been previously used to describe the specific-heat data of \SrCrO.\cite{Wang11}
The resulting contribution to the specific heat shown
as a solid line in Fig.~\ref{Fig:SpecificHeat}(a) has been
obtained with the Debye and Einstein temperatures
$\theta_{D}=123$~K, $\theta_{E1}=137$~K, $\theta_{E2}=265$~K,
$\theta_{E3}=658$~K, and $\theta_{E4}=1270$~K in agreement with
the frequency ranges where IR active phonons of the hexagonal
structure occur. The fitting curve agrees well with the specific-heat data
above $T_{JT}$, while below $T_{JT}$ it deviates from the experimental data
indicating that the magnetic and orbital contributions to the specific heat are important
below the Jahn-Teller transition. Since \BaCrO~has larger molecular mass than \SrCrO, the obtained Debye and Einstein temperatures of \BaCrO~should be lower than those of \SrCrO.\cite{Wang11} The fitted Debye-temperature ratio between \BaCrO~and \SrCrO~is 0.91, which is consistent with the ratio $\sqrt{m_{SCO}}:\sqrt{m_{BCO}}=0.88$, where $m_{SCO}$ and $m_{BCO}$ are the molecular masses of \SrCrO~and \BaCrO, respectively.

The residual specific heat divided by temperature $(C-C_{latt})/T$ is shown in
Fig.~\ref{Fig:SpecificHeat}(b). One can clearly recognize a
$\lambda$-shaped anomaly at the Jahn-Teller transition around 70~K and the magnetic anomaly below 20~K, which are well separated in
the temperature scale. The corresponding entropy $S(T)=\int_0^T
d\vartheta(C-C_{latt})/\vartheta$ reaches a value slightly lower than the expected
$S=4R\ln2$ for the sum of magnetic and orbital contribution as shown in Fig.~\ref{Fig:SpecificHeat}(c), where $R$ is the molar gas constant.
For a spin-1/2 dimer, the entropy increases by $2\ln2$ from the ordered to the disordered spin state. Because there is essentially one spin dimer per formula unit of \BaCrO, the magnetic contribution to the entropy should be $2R\ln2$. This is reached at 20~K indicated in Fig.~\ref{Fig:SpecificHeat}(c). At this temperature, the specific heat measured at different fields begins to deviate from each other on cooling. Above the Jahn-Teller transition the \emph{e} orbitals of each Cr$^{5+}$ ion are degenerate and the orbital degrees of freedom will contribute to the entropy by another $2R\ln2$, where the two Cr ions per formula unit have been taken into account.

In order to describe the magnetic contribution to the specific heat, we consider a model with the intra-dimer exchange interaction and the Zeeman interaction but ignore the inter-dimer exchange interactions which are much smaller than the intra-dimer interaction.\cite{Kofu09a} Thus, we can model the magnetic contribution to the specific heat by $C_{mag}(T)=N\frac{\partial
E}{\partial T}$ using $E=\frac{1}{Z}\sum_{i=0}^3 \epsilon_i
e^{-\beta \epsilon{_i}}$ with the partition function $Z=\sum_{i=0}^3
e^{-\beta \epsilon{_i}}$ and the energies
$\epsilon_{0,1,2,3}=0,J_0-g\mu_BH,J_0,J_0+g\mu_BH$, where $\mu_B$ is the Bohr magneton, $H$ is the magnetic field, and $\beta\equiv 1/k_BT$ with the Boltzmann constant $k_B$. The solid lines in Fig.~\ref{Fig:SpecificHeat}(d) are the calculated specific heat curves according to this model, where the intra-dimer exchange interaction $J_0=2.38$~meV was determined by inelastic neutron experiments,\cite{Kofu09a} and the \emph{g}-factor $g=1.94$ was measured by electron spin resonance experiments.\cite{Kofu09a} This simple model captures the most important features of the experimental data. The modeled specific heat reaches the maximum nearly at the same temperature as that of the experimental data. The maximum shifts to lower temperature and becomes smaller with the increase of magnetic field, since the gap between the lower-lying triplet state and the singlet state is reduced due to Zeeman interaction with the magnetic field. Because inter-dimer interactions are not considered in the model, the modeled specific heat should be smaller than the experimental data as observed in Fig.~\ref{Fig:SpecificHeat}(d). Note that this discrepancy could also originate to a certain amount from the uncertainties of the changes in the phonon density of states in the low temperature structure.

\section{Conclusion}
In summary, the infrared phonon spectrum measured with the
polarization perpendicular to the hexagonal $c_h$-axis in
\BaCrO~exhibits all expected 7$E_u$ modes above the structural
phase transition at 70~K. Below this transition, which is associated
with a Jahn-Teller distortion and orbital ordering, a splitting of
some phonon modes and the appearance of new modes indicate the
reduced symmetry in the distorted phase. In the specific heat a
clear anomaly is visible at 70~K, which does not change with applied
magnetic field. An additional low-temperature anomaly in the
specific heat is assigned to originate from magnetic excitations and
its dependence on magnetic field can be explained by considering the
thermal population of triplet states and their Zeeman splitting in
external magnetic fields. The entropy change associated with orbital
ordering is found to be reached just above the structural
Jahn-Teller transition without any sign of orbital fluctuations
below 70~K. In comparison with the isostructural compound
\SrCrO, where the polar phonons and the specific heat bear signatures
of strong fluctuations and competing interactions, the fact that this
is not the case for \BaCrO~might help to understand which interactions
are responsible for the curious behavior of \SrCrO~and single out the
differences in these otherwise very similar systems.

\begin{acknowledgments}
We want to thank V. Tsurkan and D. Vieweg for experimental support. This work is
supported by the Deutsche Forschungsgemeinschaft (DFG) partially within the
Transregional Collaborative Research Center TRR 80 (Augsburg-Munich) and
the Research Unit FOR 960. ZW and MS acknowledge the support from UA through the TRR80
graduate school. Work at UVa was supported by the U.S. Department of Energy,
Office of Basic Energy Sciences, Division of Materials Sciences and Engineering under Award DE-FG02-10ER46384.
\end{acknowledgments}

\end{document}